\documentclass[conference]{IEEEtran}
\IEEEoverridecommandlockouts
\usepackage{cite}
\usepackage{amsmath,amssymb,amsfonts}
\usepackage{algorithm}
\usepackage{algpseudocode}
\usepackage{graphicx}
\usepackage{psfrag}
\graphicspath{{./images/}}
\DeclareGraphicsExtensions{.pdf,.jpeg,.png}
\usepackage{subcaption}
\usepackage{textcomp}
\usepackage{xcolor}
\usepackage{tikz}
\usetikzlibrary{arrows.meta, positioning, patterns}
\usepackage{multicol}

\newcounter{tempEquationCounter}
\newcounter{thisEquationNumber}
\makeatletter
\newcommand\fs@spaceruled{\def\@fs@cfont{\bfseries}\let\@fs@capt\floatc@ruled
  \def\@fs@pre{\vspace{0.5\baselineskip}\hrule height.8pt depth0pt \kern2pt}%
  \def\@fs@post{\kern2pt\hrule\relax}%
  \def\@fs@mid{\kern2pt\hrule\kern2pt}%
  \let\@fs@iftopcapt\iftrue}
\makeatother

\newcommand{\BS}{\boldsymbol}
\newcommand{\bm}{\mathbf}
\newcommand{\be}{\begin{equation}}
\newcommand{\ee}{\end{equation}}
\newcommand{\bea}{\begin{eqnarray}}
\newcommand{\eea}{\end{eqnarray}}
\newcommand{\bc}{{\bm c}}
\newcommand{\x}{{\bm x}}

\newcommand{\g}{{\bm g}}

\newcommand{\br}{{\bm r}}

\newcommand{\bA}{{\bm A}}
\newcommand{\bI}{{\bm I}}
\newcommand{\bR}{{\bm R}}
\newcommand{\bW}{{\bm W}}

\newcommand{\bF}{{\bf F}}
\newcommand{\bG}{{\bf G}}
\newcommand{\bD}{{\bf D}}

\newcommand{\bB}{{\bf B}}

\newcommand{\bH}{{\bf H}}

\newcommand{\bX}{{\bf X}}

\newcommand{\bJ}{{\bf J}}
\newcommand{\bZ}{{\bf Z}}

\newcommand{\bd}{{\bf d}}

\newcommand{\bPsi}{\mbox{\boldmath$\Psi$}}

\newcommand{\bXi}{\mbox{\boldmath$\Xi$}}

\newcommand{\bPhi}{\mbox{\boldmath{$\Phi$}}}

\newcommand{\bcalX}{
\boldsymbol{\mathcal{X}}}

\newcommand{\llrrparen}[1]{
  \left(\mkern-3mu\left(#1\right)\mkern-3mu\right)}

\begin{document}

\title{Synchronization and Channel Estimation of OTFS with RF Impairments
\vspace{-0.5 cm}}

\author{{\textit{(Invited Paper)}}\\\IEEEauthorblockN{Sanoopkumar P. S.\IEEEauthorrefmark{1}, Mohsen Bayat\IEEEauthorrefmark{2}, Stephen McWade\IEEEauthorrefmark{3}
 and 
Arman Farhang\IEEEauthorrefmark{3}}

\IEEEauthorblockA{\IEEEauthorrefmark{1} Department of Electronics and Communication Engineering, South Eastern Technological University, Ireland \\ \IEEEauthorrefmark{2}Department of Computing Science and Mathematics, Dundalk Institute of Technology, Ireland\\ \IEEEauthorrefmark{3}Department of Electronic and Electrical Engineering, Trinity College Dublin, Ireland \\}

\IEEEauthorblockA{
Email: Sanoopkumar.Pungayilsasindran@setu.ie, Mohsen.Bayat@dkit.ie, \{smcwade,arman.farhang\}@tcd.ie}

\thanks{This publication has emanated from research conducted as part of the MULTIPLY-6G project. MULTIPLY-6G project has received funding from the Smart Networks and Services Joint Undertaking (SNS JU) under the European Union’s Horizon Europe research and innovation programme under Grant Agreement No 101290342. This research was also supported by Research Ireland under the US-Ireland R\&D Partnership Programme Grant Numbers 21/US/3757 and 24/US/4013.}
}

\maketitle

\begin{abstract}
Orthogonal Time Frequency Space (OTFS) modulation is a promising waveform for future wireless networks. However, its resilience to RF impairments remains relatively understudied. Low-cost RF front ends are crucial for next-generation wireless systems, yet their performance is often degraded by RF impairments such as transmit IQ imbalance (IQI), phase noise (PN), timing offset (TO), and carrier frequency offset (CFO). Hence, this paper addresses the estimation and compensation of these RF impairments in OTFS systems under high mobility. A unified system model is developed that incorporates TO, CFO, PN, IQI and other channel effects into an effective channel representation. Using the pilot with cyclic prefix (PCP), a low-peak to average power ratio (PAPR) pilot suitable for OTFS, we propose a pilot-aided synchronization and estimation framework. The dual periodicity of PCP is exploited in our proposed TO estimation technique. A maximum-likelihood-based technique is also proposed to jointly estimate the CFO and effective channel using a complex exponential basis expansion model (CE-BEM). Finally, a linear detection model is formulated in the delay-Doppler domain to mitigate residual interference caused by RF impairments. 
Our simulation results corroborate the efficacy of our proposed synchronization and channel estimation techniques. 


\end{abstract}

\section{Introduction}




Orthogonal time-frequency space (OTFS) modulation has emerged as a promising candidate waveform for future wireless networks, owing to its resilience to high-Doppler channels.
OTFS spreads the delay-Doppler (DD) domain data symbols across the entire time-frequency plane, thereby achieving full diversity offered by the doubly selective wireless channels \cite{hadani_2017}. The majority of OTFS literature thus far has been on the theoretical aspects of the waveform and there is only a small body of literature that considers practical aspects such as RF impairments \cite{ Neelam_2021_ncc,Surabhi_OTFS_PN_2019}. 
Most modern wireless communication systems use simple low-cost direct conversion transceiver (DCT) architecture. However, DCTs are highly susceptible to in-phase and quadrature signal mismatch, commonly referred to as IQ imbalance (IQI) \cite{Yoshida_2009}. Additionally, low-cost radios are prone to time-varying oscillator phase noise (PN) and carrier frequency offset (CFO). These impairments further complicate timing and frequency synchronization, which is essential for reliable communication, particularly in time-varying channels.

In OTFS literature, each of the aforementioned impairments has been investigated in isolation. 
{There have been several studies which consider OTFS with IQI \cite{Neelam_2021_ncc, Neelam_2022_Wcom_lett}. However, these use a high peak to average power ratio (PAPR) impulse pilot for IQI estimation.} Meanwhile, the authors of \cite{Surabhi_OTFS_PN_2019} demonstrate resilience of OTFS to PN while considering perfect knowledge of PN at the receiver. 
A common-phase-error (CPE) estimation method for OTFS was proposed in \cite{Bello_OTFS_pn}.
However, this method performs poorly due to the high level of inter-Doppler interference.
%
In OTFS literature on synchronization, several works, e.g., \cite{Chung2024} and \cite{SLi2024}, address TO or CFO estimation separately, whereas other studies, including \cite{Das2021} and \cite{Bayat_2022}, consider joint TO and CFO estimation using an impulse pilot.
In \cite{Bayat_2023}, a low PAPR pilot with cyclic prefix (PCP), \cite{Sanoop_pcp}, is used to jointly estimate the TO and CFO.
Although these works achieve high TO and CFO estimation accuracy, they neglect the impact of IQI and PN.
However, IQI and PN distort the periodic structure of PCP and introduce additional impairments to the signal, hindering synchronization.
To the best of our knowledge, no prior work has investigated the joint estimation and compensation of synchronization errors and RF impairments in OTFS. Hence, in this paper, we propose practical synchronization and channel estimation techniques for OTFS using the low PAPR PCP pilot, while accounting for TO, transmitter IQI (T-IQI), CFO and PN.
We present a unified formulation that incorporates TO, CFO, PN, IQI and physical channel effects into an effective channel. Using this formulation and a complex exponential basis expansion model (CE-BEM)  to approximate the effective channel, we propose a maximum-likelihood (ML) based technique to jointly estimate the CFO and BEM coefficients.
Finally, simulation results verify the effectiveness of the proposed techniques, achieving a bit error rate (BER) close to that of a system with perfect synchronisation and channel knowledge.



 \subsubsection*{Notations} Superscripts ${(\cdot)^{\rm{T}}}$ and ${(\cdot)^{\rm{H}}}$ denote transpose and Hermitian transpose, respectively. Bold lower-case characters denote vectors and bold upper-case characters denote matrices. $x[n]$ and $X[m,n]$ denote the $n$-th and $(m,n)$-th elements of $\mathbf{x}$ and $\mathbf{X}$, respectively. The function $\rm{vec}(\mathbf{X})$ vectorizes $\mathbf{X}$ by stacking its columns to form a single column vector and $\mathrm{vec}^{-1}(\mathbf{x})$ denotes the inverse operation that forms the matrix $\mathbf{X}$ from $\mathbf{x}$.
 $\mathrm{diag}(\mathbf{x})$ denotes a diagonal matrix with the elements of $\mathbf{x}$ on its main diagonal. $\mathrm{diag}(\mathbf{X})$ denotes a column vector formed by the diagonal elements of $\mathbf{X}$.
 The $p\times{p}$ identity matrix and $p \times q$ all-zero matrix are  denoted by $\mathbf{I}_p$ and $\mathbf{0}_{p\times{q}}$, respectively. The operators $\llrrparen{\cdot}_N$, $\odot$ and $\otimes$ represent Modulo$-N$, element-wise multiplication and the Kronecker product operations, respectively. The operator ${\rm rcolshift}(\mathbf{X}, l)$ denotes the matrix obtained by circularly right-shifting the columns of $\mathbf{X}$ by $l$ positions. Finally, $j = \sqrt{-1}$ represents the imaginary unit. 
\section{System Model}
\begin{figure}[t]
\begin{subfigure}{0.5\columnwidth}
   \includegraphics[width=1\columnwidth]{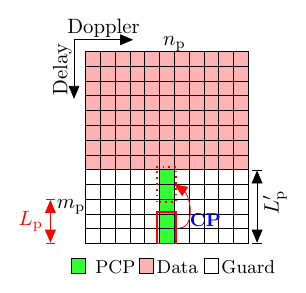}
   \caption{Transmitted Signal}
    \label{Fig:frame1}
\end{subfigure}
\hspace{-0.3 cm}
\begin{subfigure}{0.5\columnwidth}
    \includegraphics[width=1\columnwidth]{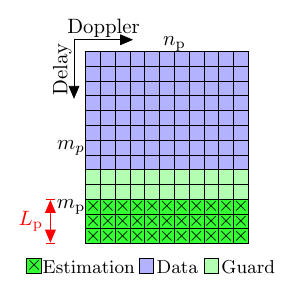}
   \caption{Received Signal}
    \label{Fig:frame2}
\end{subfigure}
\caption{The frame structure}
\label{fig:frame}
\end{figure}

In this paper, we consider a single-user OTFS system with $M$ delay bins and $N$ Doppler bins with a delay resolution of $T_{\rm s}$ seconds and Doppler resolution of $\frac{1}{MNT_{\rm s}}$ Hz, where $T_s$ is the sampling rate. At the transmitter, the information bits are mapped onto a quadrature amplitude modulation (QAM) symbol constellation and multiplexed with the pilot signal to form the DD domain symbol matrix,
$\bD \in \mathbb{C}^{M\times N}$ with the elements $D[m,n]$ for $m=0,1,\hdots M-1$ and $n=0,1, \hdots N-1$. The PCP with length $2L_{\rm p}-1$ is placed in Doppler bin $n_{\rm p}$ along the delay bins $m_{\rm p}-L_{\rm p}+1, \hdots, m_{\rm p}-1, m_{\rm p}, m_{\rm p}+1, \hdots, m_{\rm p}+L_{\rm p}-1$ for synchronisation and channel estimation \cite{Sanoop_pcp}. The DD domain frame structure is shown in Fig.~\ref{fig:frame}. 
The DD domain signal is then transformed to the delay-time (DT) domain by taking $N$-point inverse discrete Fourier transform (IDFT) across the rows of $\bD$. Thus, the DT domain transmitted signal matrix $\bX\in \mathbb{C}^{M\times N}$ can be expressed as, $\bX=\bD\bF_N^{\rm H}$.  After vectorising $\bX$ and adding a cyclic prefix (CP) of length $L_{\rm cp}$, the complex baseband signal $\x_{\rm cp}$ is obtained, which is then fed to the transmitter RF front end. The complex baseband signal equivalent to the passband signal at the output of the transmitter can be expressed as 
\begin{align}
    s(t)=\mu_{\rm T}x_{\rm cp}(t)+\nu_{\rm T}x_{\rm cp}^*(t),
\end{align}
where $x_{\rm cp}(t)$ is the complex baseband signal fed to the mixer.  The I and Q distortion parameters are given, respectively by, $\mu_{\text{T}}=\frac{1}{2}(a_{\rm {TI}}e^{j\theta_{\rm {TI}}}+a_{\rm {TQ}}e^{j\theta_{\rm {TQ}}})$ and $\nu_{\rm {T}}=\frac{1}{2}(a_{\rm {TI}}e^{j\theta_{\rm {TI}}}-a_{\rm {TQ}}e^{j\theta_{\rm {TQ}}})$ with $a_{\rm {TI}}$, $a_{\rm {TQ}}$, $\theta_{\rm {TI}}$, and $\theta_{\rm {TQ}}$ being the amplitudes and phases of I and Q branch carriers, respectively. At the receiver, the received passband signal is converted to baseband by the receive RF front end. The complex baseband signal at the output of the receiver RF front end can be expressed as,
\begin{align}\label{eqn:riqi}
    r(t)=e^{j\phi(t)+j2\pi f_{\Delta}t}y(t-t_0)+\eta(t),
\end{align}
where $\eta(t)$ is the complex additive white Gaussian noise (AWGN) with variance $\sigma^2$, $t_0$ denotes the TO between the transmitter and receiver, $\phi(t)$ is {the continuous-time phase noise at time $t$} and $f_{\Delta}$ is the CFO, which is caused by the frequency mismatch between the transmitter and receiver local oscillators \cite{Bayat_2025_2}. In (\ref{eqn:riqi}), $ y(t)=\int h_{\rm DT}(t,\tau)s(t-\tau)d\tau$ is the complex baseband signal at the input of the receiver RF front end while ignoring the AWGN
and $h_{\rm DT}(t, \tau)$ is the time-varying impulse response of the channel in DT domain.
The output of the matched filter at the receiver RF front end is sampled at intervals of $T_{\rm s}$ seconds to obtain the discrete-time signal samples, which are given by, \vspace{-0.2cm}
\begin{align} \nonumber
r[\kappa] = & ~e^{j \left(\phi[\kappa]+\frac{2 \pi \varepsilon}{MN}\kappa\right)} \sum_{\ell=0}^{L_{\rm{ch}}-1} h_{\rm DT}[\kappa,\ell] \big( \mu_{\rm T} x_{\rm cp}[\kappa-\ell-\xi] \\ \label{eqn:rec20}
&+\nu_{\rm T}x_{\rm cp}^*[\kappa-\ell-\xi] \big) + \eta[\kappa],
\end{align}
for $\kappa=0,1,\hdots, N_{\rm s}-1$ 
{where $N_{\rm s}=MN+L_{\rm cp}+\xi$ and $L_{\rm cp} \geq L_{\rm ch}-1$, with $L_{\rm ch}$ denoting the channel length.}
Additionally, $\phi[\kappa]$, $\varepsilon = \frac{f_{\Delta}}{1/(M N T_{\rm s})}$ and $\xi = \frac{t_0}{T_{\rm s}}$ denote the PN sample, the normalized CFO and TO, respectively. Furthermore, the discretized DT domain channel impulse response is represented as
\be \label{eqn:ch0} h_{\rm DT}[\kappa,\ell]=\sum_{i=0}^{P_{\rm ch}-1} \rho_i e^{j 2 \pi \nu_i (\kappa-\ell)} \delta[\ell-\ell_i], \ee
where $P_{\rm ch}$, $\rho_{i}$, $\ell_{i}$ and $\nu_{i}=\nu_{\rm max}\cos(\alpha_i)$ are the number of paths, complex gain, delay and Doppler shift of the $i$-th channel path, respectively. Additionally, $\nu_{\rm max}$ is the maximum Doppler shift and $\alpha_i$ is the angle of arrival of the $i^{\rm th}$ path. 

{
By stacking the received signal samples $r[\kappa]$ for $\kappa\!=\!0, \hdots, N_{\rm s}-1$ into the received signal vector $\br \in \mathbb{C}^{N_{\rm s} \times 1}$, \eqref{eqn:rec20} can be expressed in vector-matrix form as
\begin{align}
    \bf{r}&=\widetilde{\bPhi}\widetilde{\bPsi}\widetilde{\bXi}\widetilde{\bH}_{\rm DT}\left(\mu_{\rm T}\x_{\rm cp}+\nu_{\rm T}\x_{\rm cp}^*\right)+{ \widetilde{\BS{\eta}}}, \label{eqn:y_dt}
\end{align}
where $\widetilde{\bH}_{\rm DT}$ is the $N_{\rm w} \times N_{\rm w}$ Toeplitz matrix formed from the DT domain channel coefficients in (\ref{eqn:ch0}), with $N_{\rm w}=MN+L_{\rm cp}$.
Additionally, $\widetilde{\bXi}=[\mathbf{0}_{\xi \times N_{\rm w}},\bI_{N_{\rm w}}] \in \mathbb{R}^{N_{\rm s} \times N_{\rm w}}$ and  $\widetilde{\bPsi}={\rm diag} \big(e^{j\frac{2\pi\varepsilon (0)}{MN}}, e^{j\frac{2\pi\varepsilon(1)}{MN}}, \hdots, e^{j\frac{2\pi\varepsilon (N_{\rm s}-1)}{MN}} \big) \in \mathbb{C}^{N_{\rm s} \times N_{\rm s}}$ represent the TO and CFO matrices, respectively.
Furthermore, $\widetilde{\bPhi}={\rm diag}\big(e^{j\phi[0]}, e^{j\phi[1]}, \hdots, e^{j\phi[N_{\rm s}-1]}\big) \in \mathbb{C}^{N_{\rm s} \times N_{\rm s}}$ denotes the PN matrix and ${ \widetilde{\BS{\eta}}}$ is AWGN vector.
}

Although the effects of CFO, TO and PN can be incorporated into an effective channel and corresponding compensation methods can be applied, the presence of TO can significantly degrade system performance. Therefore, precise TO estimation is essential prior to CFO and PN estimation and compensation. Moreover, as shown in (\ref{eqn:riqi}), T-IQI introduces conjugate interference in the received signal, which is not addressed by existing synchronisation and channel estimation methods. Consequently, in the following sections, we develop a TO estimation technique, followed by a joint estimation approach for CFO, channel, PN and T-IQI.

\section{Proposed Synchronization and Channel Estimation Techniques}
As discussed earlier, RF impairments such as IQI and PN in practical OTFS systems distort the periodic structure of the PCP. This poses significant challenges to synchronization and channel estimation \cite{Bayat_2023}.
Therefore, in this section, we propose novel TO, CFO and channel estimation techniques for OTFS in presence of T-IQI and PN.

\subsection{Proposed TO Estimation Technique} 
PCP exhibits a dual periodicity property in the DT domain. The periodicity in the delay dimension is introduced by CP, whereas the periodicity in time dimension arises from spreading each Doppler domain pilot sample along time through IDFT operation. 
In our proposed time synchronization technique, we exploit this dual periodicity in both the delay and time dimensions to estimate the TO.

To construct PCP, we employ a ZC sequence, defined as $p_0[m] = \sqrt{N \sigma^2_{\rm{p}}} e^{-j \frac{\pi \mu m (m+1)}{L_{\rm{p}}}},~m = 0, \ldots, L_{\rm{p}} - 1$, where $\mu$ is the root index satisfying ${\rm gcd}(L_{\rm{p}}, \mu) = 1$ and $N \sigma^2_{\rm{p}}$ denotes the pilot power.
The last $L_{\rm{p}} - 1$ samples of this sequence are then prepended at the beginning of the sequence as a CP, denoted by $\bm{p} \in \mathbb{C}^{(2L_{\rm{p}} - 1) \times 1}$.
This pilot is inserted at the Doppler bin $n_{\rm{p}}$ and the delay bins $m_{\rm{p}} - L_{\rm{p}} + 1$ to $m_{\rm{p}} + L_{\rm{p}} - 1$ along data symbols with $m_{\rm p}\in\{L_{\rm p}-1,\ldots,M-L_{\rm p}\}$ and $n_p\in\{0,\ldots,N-1\}$, see Fig.~\ref{fig:frame}~(a).

{
After transmitting and receiving the data, including the PCP, the first $MN$ samples of the received vector ${\bm r}$ in (\ref{eqn:y_dt}) are used to construct $\bR = {\rm vec}^{-1}(\br) \in \mathbb{C}^{M \times N}$ in the DT domain.
The matrix 
$\bm{P} = \big[ \bm{0}_{N\times (m_{\rm p}-L_{\rm p}+1)},\, (\overline{\bm P}\bm F_N^{\rm H})^{\rm T},\, \bm{0}_{N\times (M-m_{\rm p}-L_{\rm p})} \big]^{\rm T}$, represents the transmitted DT pilot signal, where $\overline{\bm P} = \big[ \bm{0}_{(2L_{\rm p}-1)\times n_{\rm p}},\, \bm{p},\, \bm{0}_{(2L_{\rm p}-1)\times(N-n_{\rm p}-1)} \big]$.
The proposed technique exploits the aforementioned dual periodicity of the PCP in the DT domain to estimate the TO. To this end, the correlation between the received signal and the transmitted DT pilot signal is computed within each delay block of an OTFS frame. The resulting correlation values are then accumulated over all delay blocks in the time dimension to obtain the correlation function
\begin{equation} \label{eqn:to_corr} 
{\lambda}[m'] = \frac{1}{MN} \sum_{k=0}^{N-1} \Big| \sum_{m=0}^{M-1} R[m,k] P^*[(\!(m-m')\!)_M,k] \Big|, 
\end{equation}
where $m' = 0,\ldots,M-1$. 
}

\begin{figure}
    \centering \includegraphics[width=0.95\columnwidth]{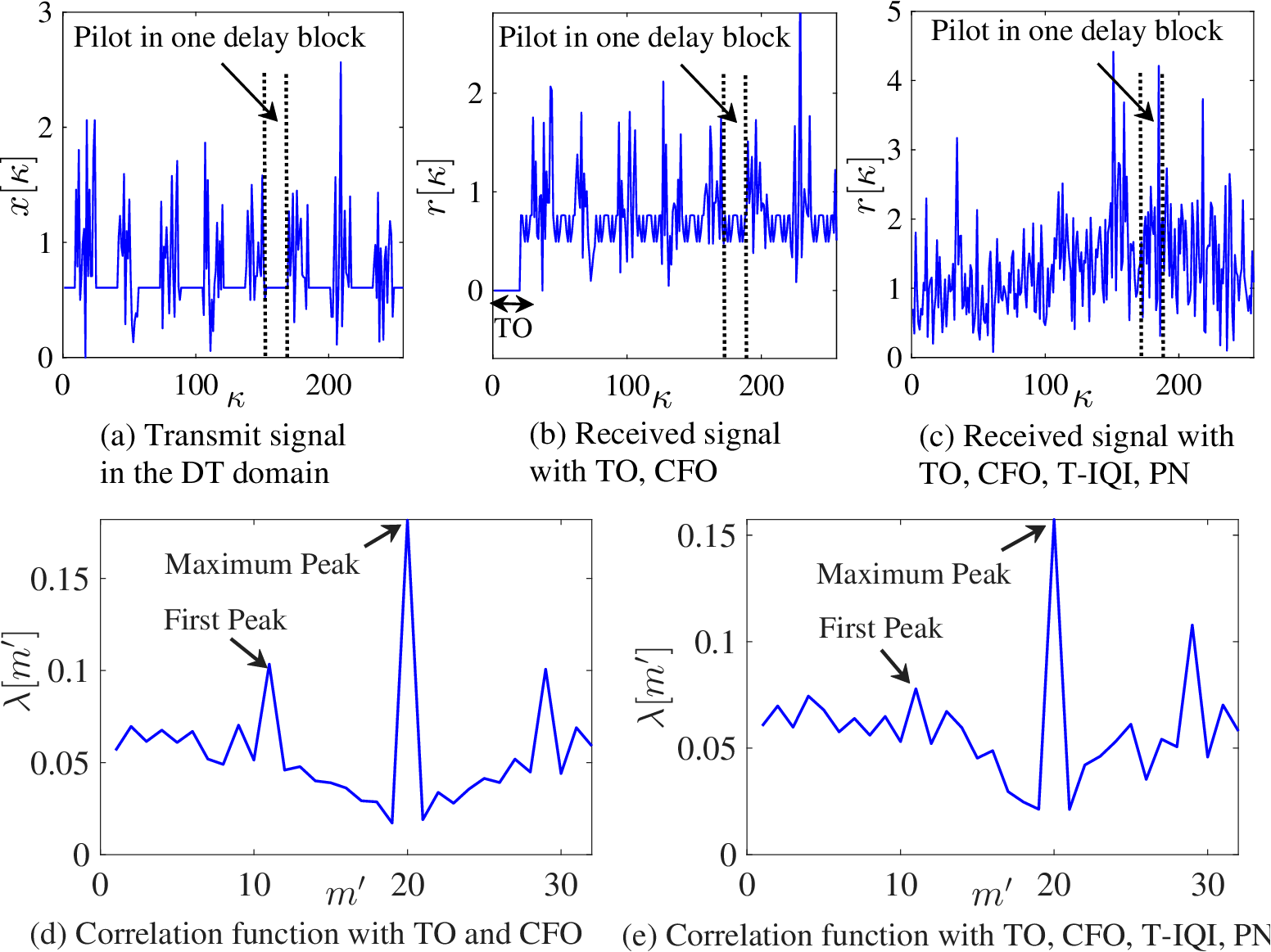}
    \caption{T-IQI, TO, CFO and PN effect on the received signal and the correlation function in the absence of AWGN}
    \label{fig:signal}
\end{figure}

As shown in (\ref{eqn:y_dt}), the matrix $\bm{R}$ represents the received signal that is affected by T-IQI, TO, CFO, and PN. These RF impairments, as illustrated in Figs.~\ref{fig:signal}(a)-(c) {for $M=64$, $N=32$, and $L_{\rm cp}=9$}, significantly disrupt the periodicity of the pilot signal across delay blocks in the DT domain. Consequently, the correlation function differs from that of an ideal OTFS system.
%
%
{The three-peak structure in Figs.~\ref{fig:signal}(d) and~\ref{fig:signal}(e) follows from the PCP construction. The maximum peak arises when the received PCP is fully aligned with its template. The two adjacent peaks, located $L_{\rm p}$ samples apart, stem from partial overlapping of the template with the received pilot in the CP part. 
Under T-IQI and PN, all peaks are attenuated, but the first peak is the most affected and, as labeled in Fig.~\ref{fig:signal}(e), may become indistinguishable from the side peaks.}
Therefore, the first peak, as proposed in \cite{Farhang_2024}, does not provide a reliable estimation accuracy in practical systems under RF impairments.
{In contrast, since the maximum peak accumulates the largest correlation energy, it remains dominant and the most robust to the RF impairments.}
%
Hence, we propose to estimate the TO by detecting the maximum peak, whose lag corresponds to the received PCP displacement, as follows 
\begin{equation} \label{eqn:highest_peak} 
\widehat{\xi} = \big(\!\big( \arg\max_{m'} \big\{ |{\lambda}[m']| \big\} - L_{\rm cp} \big)\!\big)_M.
\end{equation}
Since the pilot position $m_{\mathrm{p}}$ is embedded in the template $P[(\!(m-m')\!),k]$, the lag index $m'$ represents the displacement of the received PCP relative to its transmitted position. This displacement equals the TO plus the CP length $L_{\rm cp}$, since the CP precedes the frame within the correlation window.
Hence, the correlation metric in (\ref{eqn:to_corr}) is maximized at $m'\!=\!(\!(\xi+L_{\rm cp})\!)_{M}$, where the received PCP aligns with its template. Since $0 \leq \xi \leq M-1$, this mapping is one-to-one and subtracting $L_{\rm cp}$ while taking modulo $M$, as in (\ref{eqn:highest_peak}), uniquely recovers the TO.

%
Besides degrading TO estimation, RF impairments also affect CFO and channel estimation. Therefore, the next subsection proposes a joint CFO and channel estimation technique for OTFS systems under T-IQI and PN.

\subsection{Joint CFO and Effective Channel Estimation Technique}
{
After correcting the TO and removing CP, the received signal can be expressed as, 
\begin{align}
     \overline{\bf{r}} &=\bPhi\bPsi\bH_{\rm DT}\left(\mu_{\rm T}\x+\nu_{\rm T}\x^*\right) +{ \BS{\eta}}.
     \label{eqn:y_dt_rcp}
\end{align}
where, 
${\bPsi}=e^{j\frac{2\pi\varepsilon (\widehat{\xi}+L_{\rm cp})}{MN}}{\rm diag} \big(1, e^{j\frac{2\pi\varepsilon(1)}{MN}}, \hdots, e^{j\frac{2\pi\varepsilon (MN-1)}{MN}} \big)$
and $\widetilde{\bPhi}\!=\!{\rm diag}\big(e^{j\phi[\widehat{\xi}+L_{\rm cp}]}, 
\hdots, e^{j\phi[\widehat{\xi}+L_{\rm cp}+MN-1]}\big)$. In addition, the channel matrix is $\bH_{\rm DT}=\bR_{\rm cp}\widetilde{\bH}_{\rm DT}\bA_{\rm cp}$, and $\BS{\eta}=\bR_{\rm cp}\widetilde{\BS{\eta}}$. Here, $\bA_{\rm cp}=[\bJ_{\rm cp}^{\rm T},\bI_{MN}^{\rm T}]^{\rm T} \in \mathbb{R}^{N_{\rm w} \times MN}$ is the CP addition matrix, $\bR_{\rm cp}=[\mathbf{0}_{MN \times L_{\rm cp}},\bI_{MN}]$ is the CP removal matrix and $\bJ_{\rm cp}$ is formed by the last $L_{\rm cp}$ rows of $\bI_{MN}$.
The received signal in DD domain, $\bZ$, is obtained by taking the discrete Zak transform (DZT) of $\overline{\br}$, i.e., $\bZ=\overline{\bR}\bF_{\rm N}$, where $\overline{\bR}=\rm{vec}^{-1}(\overline{\br})$.

The DT domain received signal in (\ref{eqn:y_dt_rcp}) can be expressed as,
\begin{align}
    \overline{\bf{r}}&=\bPsi\left(\bf{G}_{\rm DT,1}\x+\bf{G}_{\rm DT,2}\x^*\right)+{ \BS{\eta}}\nonumber\\ &=\overline{\BS{\mathcal{X}}}\mathfrak{h}_1+\overline{\BS{\mathcal{X}}}^*\mathfrak{h}_2+{ \BS{\eta}}, \label{eqn:r_dt_bem}
\end{align}
where ${\bf{G}}_{\rm DT,1}\!\!=\!\!\mu_{\rm T} \bPhi \bH_{\rm DT}$, ${\bf{G}}_{\rm DT,2}\!\!=\!\!\nu_{\rm T}\bPhi\bH_{\rm DT}$. Considering $\g^p_{{\rm{DT}},l}\!\!=\!\!{\rm diag}\left({\rm rcolshift}\left( \bG_{{\rm DT},p},l\right)\right)$ for $p=1,2$ and $l=0,\ldots,L-1$, 
$\mathfrak{h}_p = [ (\g^p_{{\rm{DT}},0})^{\rm T}, (\g^p_{{\rm{DT}},1})^{\rm T}, \ldots,(\g^p_{{\rm{DT}},L-1} )^{\rm T} ]^{\rm T}$
and $\overline{\BS{\mathcal{X}}}=\left[\BS{\mathcal{X}}_{0}, \BS{\mathcal{X}}_{1}, \hdots, \BS{\mathcal{X}}_{L-1} \right]$ 
where  $\BS{\mathcal{X}}_{l}={\rm{diag}}\left(x[\llrrparen{0-l}_N],x[\llrrparen{1-l}_N], \hdots, x[\llrrparen{MN-1-l}_N]\right)$. 
Using CE-BEM with $Q$ basis coefficients and an oversampling factor $K$, we can approximate $\g^p_{{\rm{DT}},l}$ as $\g^p_{{\rm{DT}},l}\approx\bB\bc_{p,l}$, where the basis function $\mathbf{B}$ is defined by ${B}[\kappa,q]=e^{\frac{j2\pi(q-Q/2)\kappa}{KMN}}$, for $q= 0,1, \hdots, Q-1$ and $\kappa = L_{\rm cp}\!+\!0, L_{\rm cp}\!+\!1, \!\hdots, N_{\rm s}$, and $\bc_{p,l}\in \mathbb{C}^{Q\times 1}$ denotes the basis coefficient vector corresponding to the $l$-th channel tap. Hence, (\ref{eqn:r_dt_bem}) is reformulated as,
\begin{align} \overline{\br}&\approx\overline{\BS{\mathcal{X}}}\left(\bI_L\otimes \bB \right)\bc_1+\overline{\BS{\mathcal{X}}}^*\left(\bI_L\otimes \bB \right)\bc_2+{ \BS{\eta}} =\boldsymbol{\mathcal{X}}\bc+{ \BS{\eta}}, \label{eqn:rdt_bem_3}
\end{align}
where $\bc_p=[\bc_{p,0}^{\rm T}, \ldots, \bc_{p,L-1}^{\rm T}]^{\rm T}$ and $\bc=[\bc_1^{\rm T}, \bc_2^{\rm T}]^{\rm T}$.
At the receiver, the received samples with time indices $\Gamma_{\rm p}^{\rm Est}=\left\lbrace L_{\rm cp}\!+\!{n}m_{\rm p}, L_{\rm cp}+nm_{\rm p}\!+\!1, \hdots, L_{\rm cp}\!+\!nm_{\rm p}\!+\!L_{\rm p}\!-\!1\right\rbrace$, $n \!=\! 0,1, \hdots, N-1$, are used for pilot aided channel estimation \cite{Sanoop_pcp}. Thus, the the received pilot vector, $\br_{\rm p}$ is formed by stacking the samples $\overline{r}[n],\, \forall n\in \Gamma_p^{\rm Est}$. Using the CE-BEM approximation of the effective channel $\br_{\rm p}$ is expresssed as, 
\begin{align} \label{eqn:P_rec2} 
{\br}_{\rm p} \!\!&=\! {\bPsi_{{\rm p},\varepsilon}}\bcalX_{\rm p}(\bI_N \!\otimes\! (\bI_{L_{\rm p}} \!\!\otimes\! \bB_{\rm p}))\bc \!+\! {\boldsymbol{\eta}_{\rm p}} \!=\! {\bPsi_{{\rm p},\varepsilon}} \overline{\bcalX}_{\rm p}\bc + {\boldsymbol{\eta}_{\rm p}},
\end{align} 
where
$\bcalX_{\rm p}\!=\![(\bcalX_{\rm p,0})^{\rm{T}},(\bcalX_{\rm p,1})^{\rm{T}},\ldots,(\bcalX_{{\rm p},N-1})^{\rm{T}}]^{\rm{T}}$, $\bcalX_{\rm p, n}={\rm rcolshift}\left({\rm{diag}}\left(x[nm_{\rm p}],x[nm_{\rm p}\!+\!1], \hdots, x[nm_{\rm p}+L_{\rm p}\!-\!1]\right), l\right)$, $\bB_{\rm p} \in \mathbb{C}^{L_{\rm p}\times Q}$ denotes the submatrix formed by stacking the rows of $\bB$ with the indices in $\Gamma_{\rm p}^{\rm Est}$, $\bPsi_{{\rm p},\varepsilon}={\rm diag}\left(e^{j2\pi\frac{\varepsilon \kappa}{MN}}\right), \, \forall \kappa\in \Gamma_{\rm p}^{\rm Est}$ and $\overline{\bcalX}_{\rm p}=\bcalX_{\rm p}(\bI_N \!\otimes\! (\bI_{L_{\rm p}} \!\!\otimes\! \bB_{\rm p}))$. Furthermore, ${\boldsymbol{\eta}_{\rm p}}$ includes AWGN noise samples at the received pilot indices on its entries. From (\ref{eqn:P_rec2}), the ML joint estimation of $\varepsilon$ and $\bc$ can be formulated as $    (\widehat{\bc}, \widehat{\varepsilon}) = \arg \max_{{\bc},{\varepsilon}} \left\lbrace {\rm{ln}} \big(f({\br};{\bc},{\varepsilon})\big) \right\rbrace$,
where $f({\br};{\bc},{\varepsilon})$ is the probabilty density function of $\br_{\rm p}$ conditioned on $\varepsilon$ and $\bc$, 
\begin{align}\label{eqn:ML}
   f({\br_{\rm p}};{\bc},{\varepsilon}) \!=\! \frac{1}{\big(\pi \sigma_{\rm{\eta}}^2\big)^{NL_{\rm{p}}}}e^{-\frac{1}{\sigma_{\rm{\eta}}^2}{\left[{\br}-\bPsi_{{\rm p},\varepsilon}\overline{\bcalX}_{\rm p} {\bc}\right]^{\rm H} \left[{\br}-\bPsi_{{\rm p},\varepsilon}\overline{\bcalX}_{\rm p} {\bc}\right]
   }}.\!\!
\end{align}
Using (\ref{eqn:ML}) and removing the constant terms independent of the parameters to be estimated, the estimation problem can be reformulated as, 
\begin{align}\label{eqn:ml_cost}
    (\widehat{\bc},\widehat{\varepsilon}) \!=\! \arg \max_{{\bc},{\varepsilon}}
\!\Big\lbrace\!-\frac{1}{\sigma_{\rm{\eta}}^2} [{\br_{\rm p}}\!-\!{\bPsi_{{\rm p},\varepsilon}} \overline{\bcalX}_{\rm p} {\bc}]^{\rm{H}} [{\br_{\rm p}}\!-\!{\bPsi_{{\rm p},\varepsilon}} \overline{\bcalX}_{\rm p} {\bc}] \!\Big\rbrace\!.\!
\end{align}
Let $\varepsilon'$ be the value of $\varepsilon$ which maximises the cost function in (\ref{eqn:ml_cost}). Since, $\bc$ and $\varepsilon$ are independent variables, $\widehat{\bc}({\varepsilon}')$ can be obtained as 
\be
{\widehat{\bm{c}}}({\varepsilon}') = \overline{\bcalX}_{\rm p}^{\dag} \bPsi_{{\rm p},\varepsilon'}^{\rm{H}} {\br}_{\rm p},
\label{eqn:c_est} 
\ee
where $\overline{\bcalX}_{\rm p}^{\dag}=\big( \overline{\bcalX}_{\rm p}^{\rm{H}} \overline{\bcalX}_{\rm p} \big)^{-1} \overline{\bcalX}_{\rm p}^{\rm{H}}$. 
Substituting, (\ref{eqn:c_est}) back to (\ref{eqn:ml_cost}), the CFO can be estimated as 
\begin{align}
\widehat{\varepsilon}=\arg \max_{{\varepsilon}}~g({\varepsilon}),  
\end{align}
where $g({\varepsilon}) = {\br}_{\rm p}^{\rm{H}} \bPsi_{{\rm p},\varepsilon} \overline{\bcalX}_{\rm p} \overline{\bcalX}_{\rm p}^{\dag} \bPsi_{{\rm p},\varepsilon}^{\rm{H}} {\br}_{\rm p}.$
Using the estimated CFO, the BEM coefficients are obtained from \eqref{eqn:c_est}. 
\subsection{Equalization and Data Detection}
The received signal after compensating the TO and CFO can be expressed in DD domain as, 
\begin{align}\label{eqn:zg}
{\bf{z}}={\bf{G}}_{\rm DD,1}\bd+{\bf{G}}_{\rm DD,2}(\bd^{\#})^{*}+{ \BS{\eta}}_{\rm DD}, 
\end{align}
where 
${\bf{G}}_{\rm DD,1}=(\bF_N\otimes \bI_M) {\bf{G}}_{\rm DT,1}({\bF}_N^{\rm H}\otimes {\bI}_M)$ and ${\bf{G}}_{\rm DD,2}=(\bF_N\otimes \bI_M){\bf{G}}_{\rm DT,2} (\bF_N^{\rm H}\otimes \bI_M)$. In (\ref{eqn:zg}), $\bd^{\#}={\rm vec}(\bD^{\#})$ where $D^{\#}[m,n]=D[m,\llrrparen{N-n}_{N}]$ for $m=0,1,\hdots, M-1$ and $n=0,1,\hdots, N-1$. Thus,  each transmit symbol experiences interference from the symbol in the same delay bin and the corresponding mirror Doppler bin. This IDI is called mirror Doppler interference (MDI).  
Due to MDI, the received signal is widely linear as it is a linear combination of $\bd$ and $(\bd^{\#})^{*}$. Hence, widely linear filtering methods can be used for compensating IQI \cite{Yoshida_2009}. In particular, we can construct a new signal vector, $\overline{\bf{z}}=[\bf{z}^{\rm T},(\bf{z}^*)^{\rm T}]^{\rm T}$, given by 
\begin{align}\label{eqn:zbar}
\overline{\bf{z}}&={\bW}\overline{\bf{d}}+\overline{{ \BS{\eta}}}_{\rm DD}, 
\end{align}
where, ${\bW}=\begin{bmatrix}
        \bG_{\rm DD,1} & \bG_{\rm DD,2}\\
        (\bG_{\rm DD,2}^{\#})^* & (\bG_{\rm DD,1}^{\#})^* 
    \end{bmatrix}$, $\overline{\bf{d}}=[\bf{d}^{\rm T},(\bf{d}^*)^{\rm T}]^{\rm T}$ and $\overline{{ \BS{\eta}}}_{\rm DD}=[{ \BS{\eta}}_{\rm DD}^{\rm T},({ \BS{\eta}}_{\rm DD}^*)^{\rm T}]^{\rm T}$.
Then, we can use (\ref{eqn:zbar}) for data detection using linear techniques like minimum mean square error (MMSE) or iterative techniques \cite{lsmr_sic}. 
\section{Numerical results}
\begin{figure*}[!t]
\begin{multicols}{4}
\centering 
    \includegraphics[scale=0.25]{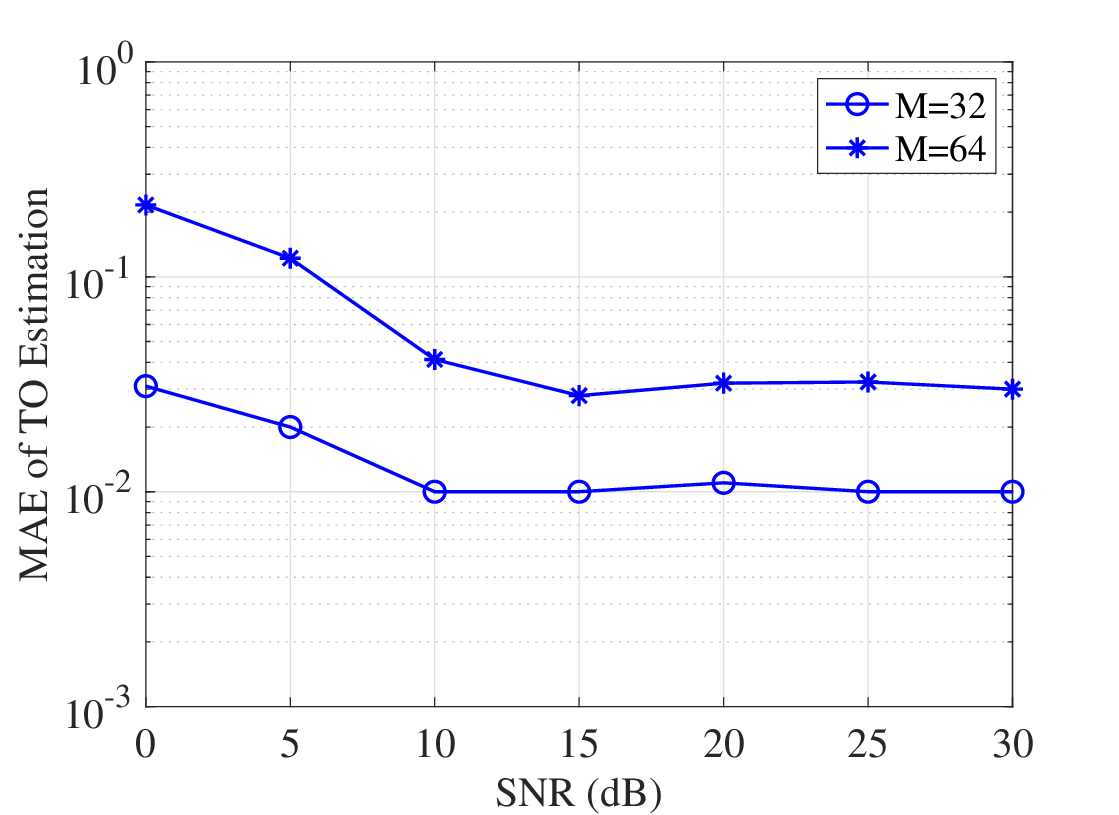}\par \vspace{-0.2cm} 
    \caption{TO estimation MAE}
    \label{fig:mae_to}
    \includegraphics[scale=0.25]{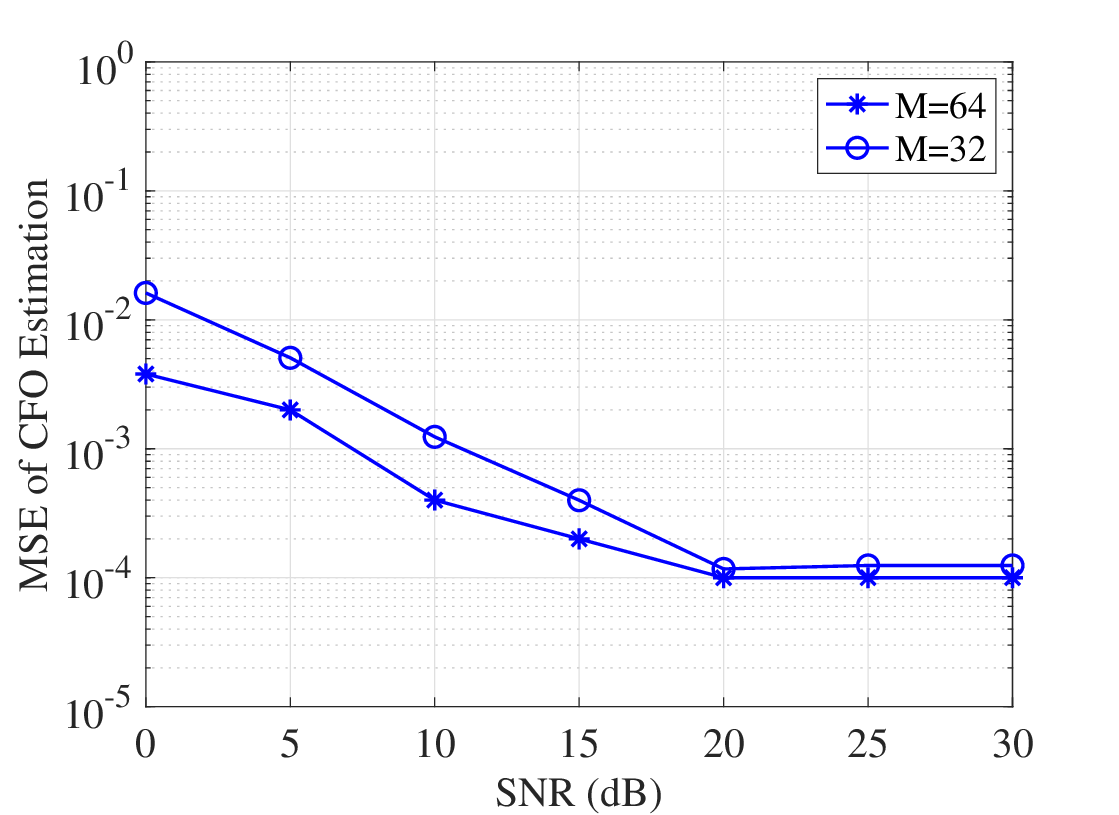}\par \vspace{-0.2cm} 
    \caption{CFO estimation MSE}
    \label{fig:nmse_cfo}
    \includegraphics[scale=0.25]{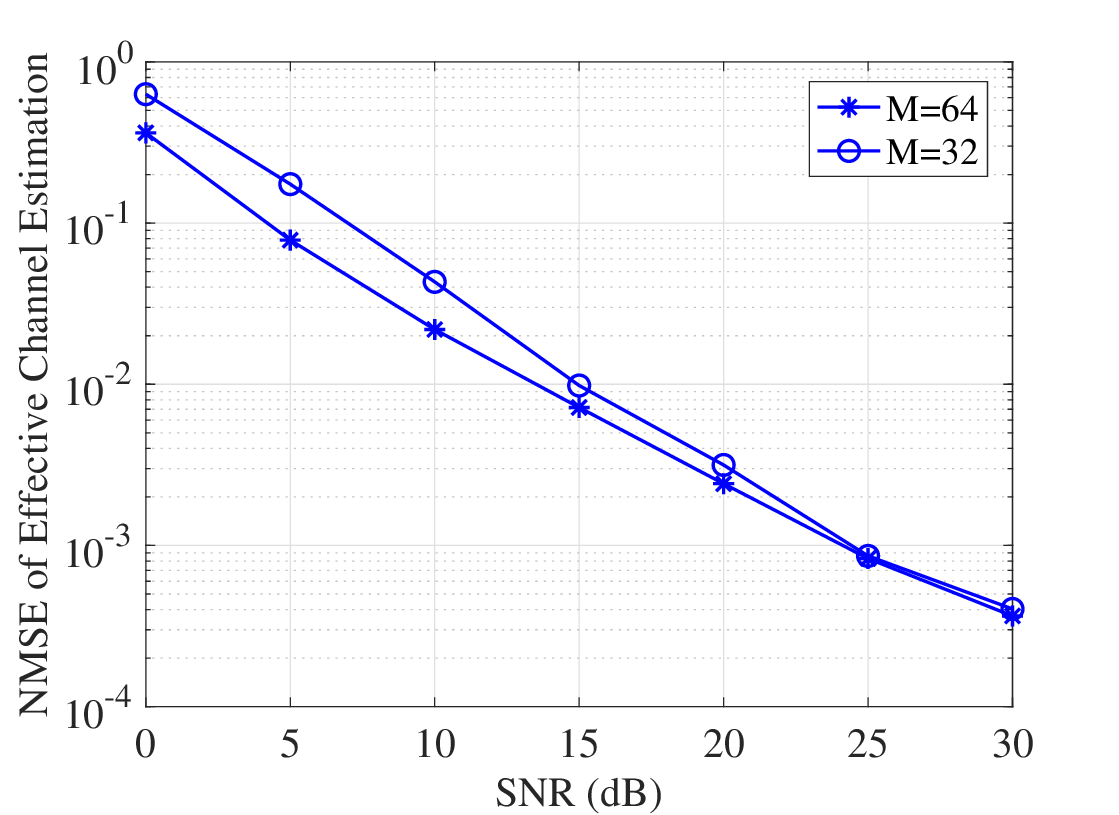}\par \vspace{-0.2cm}
    \caption{Channel NMSE}
    \label{fig:nmse_channel}
    \includegraphics[scale=0.25]{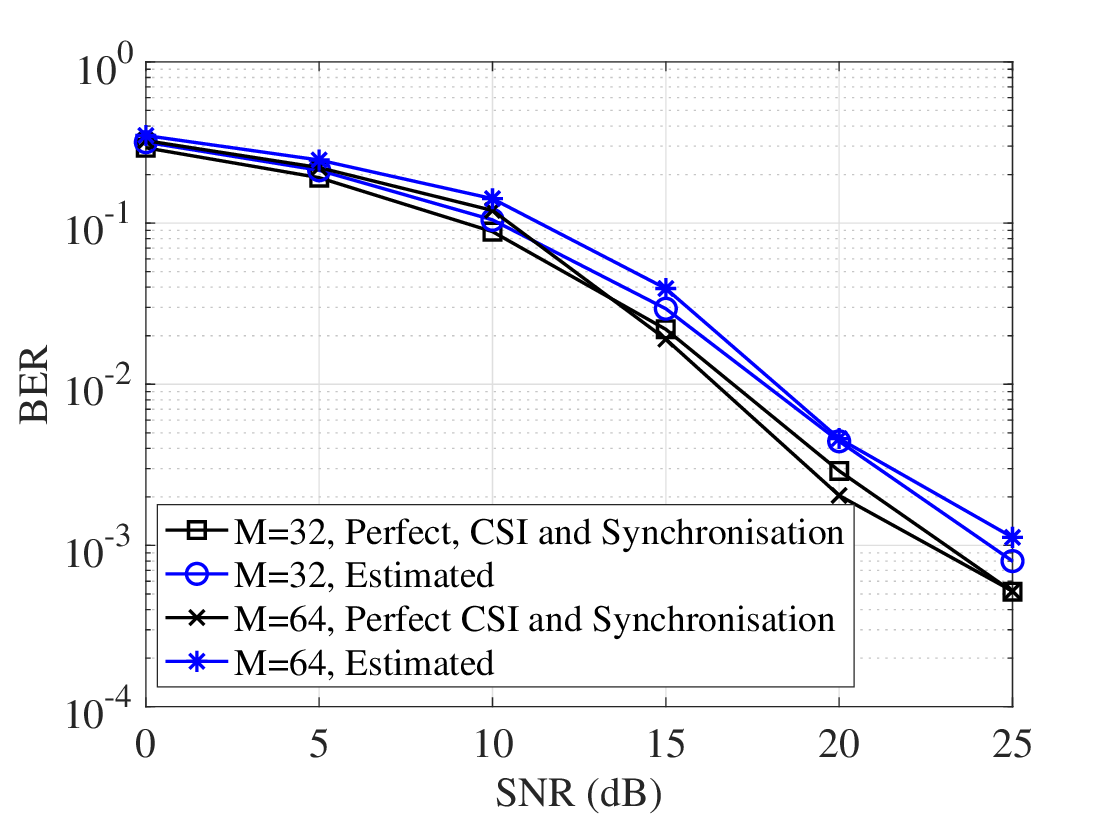}\par \vspace{-0.2cm} 
    \caption{BER performance}
    \label{fig:ber_mmse}
\end{multicols}
\end{figure*}
In this section, we numerically analyze the performance of the proposed estimation techniques under RF impairments. 
Simulations were carried out using a sampling period of $1.0417$~$\mu {\rm s}$, a carrier frequency of $5.9$~GHz, $M\in\{32,64\}$ delay bins and $N=16$ Doppler bins per frame. 
The PCP spans the delay dimension over $m_{\rm p}=M-2L_{\rm p}+1, \ldots, M-1$ at $n_{\rm p} = 1$, with $L_{\rm p}=7$ and the pilot power of $30$~dB.
The PN is generated as a first-order autoregressive process with increment variance $4\pi\beta_{\rm 3dB}T_{\rm s}$, where $\beta_{\rm 3dB}=400$~Hz is the $3$~dB bandwidth \cite{3gpp}. The amplitude imbalance, phase imbalance, TO and CFO, were uniformly distributed within the ranges $[0, 2]$~dB, $[0, 5]^\circ$, $[0, M-1]$ and $(-0.5, 0.5]$, respectively. For the BEM, $Q=7$ and $K=2$ were used. We employ the extended vehicular~A (EVA) channel model, \cite{3gpp}, with Doppler spread of $2.7$~kHz. 

Fig.~\ref{fig:mae_to} shows the mean absolute error (MAE) of the TO estimation error versus the signal-to-noise ratio (SNR) for two different OTFS frame sizes with $M = \{32, 64\}$. 
As observed, the proposed TO estimation technique, based on detecting the maximum peak of the correlation function, achieves an accuracy with an error of less than one sample for all SNRs, which can be effectively absorbed by the CP. The results also indicate that increasing the number of delay bins deteriorates the accuracy of the TO estimates. Increasing $M$ increases the spacing between pilot samples in different delay blocks in the time domain, thereby exposing the pilot to rapid time variation of the channel. This degrades the TO estimation accuracy.  

The mean squared error (MSE) of CFO estimation is shown in Fig.~\ref{fig:nmse_cfo}. It can be observed from the figure that the proposed CFO estimation provides an accurate CFO estimate with a mean squared error in the order of $10^{-4}$. Fig.~\ref{fig:nmse_channel} shows the normalized MSE (NMSE) of the proposed channel estimation technique. 
Both Figs.~\ref{fig:nmse_cfo} and \ref{fig:nmse_channel} show that increasing the number of delay bins leads to more accurate estimates of the CFO and the effective channel. This improvement is attributed to the higher Doppler resolution achieved as $M$ increases.
%

Finally, we have evaluated the effectiveness of the proposed synchronization and joint estimation techniques, as well as the equalization method based on the widely linear system model in (\ref{eqn:zbar}). We used the equalization method in \cite{lsmr_sic} and evaluated the BER performance as shown in Fig.~\ref{fig:ber_mmse}. The results show that the proposed synchronization and joint estimation techniques incur a less than a 2~dB loss compared to the perfect-CSI case.




\section{Conclusion}
In this paper, we investigated the joint estimation and compensation of RF impairments, namely T-IQI, PN, TO and CFO, in OTFS systems under high mobility. We developed a comprehensive system model that combines the effects of PN, channel and IQI into an effective channel. Furthermore, we used PCP to develop pilot-aided TO synchronization, joint CFO and effective channel estimation. The proposed TO estimation technique leverages the dual periodicity of the PCP to develop a correlation metric that provides an accurate estimate even in the presence of RF impairments. Moreover, the effective channel coefficients were approximated using CE-BEM to develop a joint CFO and channel estimation technique based on the  maximum-likelihood criterion. Finally, a widely linear system model was developed for the received signal in the DD domain to compensate for interference caused by RF impairments. The presented BER and estimation performance analyses demonstrate that the proposed method is suitable for practical low-cost radios in high-mobility scenarios.

\bibliographystyle{IEEEtran}
\bibliography{biblio}

\end{document}